\documentclass[prb,twocolumn]{revtex4}%

\usepackage{graphicx}
\usepackage{amsmath}
\usepackage{ulem}
\usepackage{color}

\newcommand{\be}{\begin{equation}}
\newcommand{\ee}{\end{equation}}
\newcommand{\bea}{\begin{eqnarray*}}
\newcommand{\eea}{\end{eqnarray*}}
\newcommand{\bean}{\begin{eqnarray}}
\newcommand{\eean}{\end{eqnarray}}

\begin{document}

\draft
\title
{\bf Thermoelectric and electron heat rectification properties of
quantum dot superlattice nanowire arrays}

\author{David M T Kuo}

%\author{David M T Kuo}
\address{Department of Electrical Engineering and Department of Physics, National Central
University, Chungli, 320 Taiwan}

%\address{$^{2}$Research Center for Applied Sciences, Academic Sinica,
%Taipei, 11529 Taiwan}

%\affiliation{$^3$ Department of Physics, National Cheng Kung
%University, Tainan, 701 Taiwan}

\date{\today}

\begin{abstract}
Heat engines made of quantum-dot (QD) superlattice nanowires
(SLNWs) offer promising applications in energy harvesting due to
the reduction of phonon thermal conductivity. In solid state
electrical generators (refrigerators), one needs to generate
(remove) large amount of charge current (heat current).
Consequently, a high QD SLNW density is required for realistic
applications. This study theoretically investigated the properties
of power factor and electron heat rectification for an SLNW array
under the transition from a one dimensional system to a two
dimensional system. The SLNW arrays show the functionality of heat
diodes, which is mainly attributed to a transmission coefficient
with a temperature-bias direction dependent characteristic.
\end{abstract}

\maketitle

\section{Introduction}
The semiconductor quantum dots (QDs) resulting from the quantum
confinement of heterostructures exhibit atom-like discrete
electron energy levels, QDs are also called artificial atoms. Due
to the localized wave functions of nanoscale QDs, electron Coulomb
interactions are too strong to be ignored. The  Coulomb blockade
effect [\onlinecite{Su},\onlinecite{Field}] and the Kondo effect
[\onlinecite{Madhavan},\onlinecite{Cron}] are experimentally
reported to reveal how electron Coulomb interactions influence
electron transport in different temperature regimes. Because the
size and location of individual QDs can be precisely controlled by
the modern semiconductor technique,  the sophisticated QD molecule
junction systems can be laid out
[\onlinecite{Deng}-\onlinecite{Gustavsson}]. Recently, nanowires
with end QDs are proposed to clarify the Majorana bound state,
which is believed to be very useful in the application of quantum
computing.[\onlinecite{Deng}] Based on the charge filter feature
of QDs, the transport behavior of single electron transistors made
of QDs of different materials have been extensively
studied.[\onlinecite{Guo}-\onlinecite{Kuba}]. In addition, single
photon sources[\onlinecite{Michler}-\onlinecite{Chang}] and single
photon detectors[\onlinecite{Gustavsson}] made of QDs are proposed
for the applications of quantum optics.

Apart from the above promising applications, scientists have also
focused on the thermoelectric (TE) properties of QD 3-D and 2-D
crystals for the applications of energy
harvesting[\onlinecite{Harman},\onlinecite{Chen}]. The figure of
merit $ZT=2$ of QD 3-D superlattice was experimentally reported
[\onlinecite{Harman}]. This enhancement is due to the reduction of
phonon thermal conductivity, which is mainly attributed to the
increase of phonon scattering resulting from the interfaces of
QDs. $ZT$ values  higher than 4 and 6 are respectively predicted
for 5~nm diameter $PbSe/PbS$ and $PbTe/PbSe$ of superlattice
nanowires (SLNW) at $77K$ in Ref. [\onlinecite{Lin}], where the
free electron model is employed to illustrate the electron
thermoelectric properties. Conventional thermoelectric materials
employ the doping method to provide the carriers
.[\onlinecite{Chen},\onlinecite{Lin}] However, Mahan and Wood
proposed to utilize the thermionic procedure to provide the
carriers [\onlinecite{Mahan}] while avoiding the electronic
defects caused by ion implantation.[\onlinecite{Rama}]

The phonon thermal conductivity of silicon/germanium SLNWs can be
reduced one order magnitude when compared with that of silicon
nanowires.[\onlinecite{Hu}] This implies that the $ZT$ of SLNWs
has a potential to reach high
values.[\onlinecite{Karbaschi}-\onlinecite{Whitney2}] In solid
state electrical generators (refrigerators), one needs to generate
(remove) large amount of charge current (heat current).
Consequently, a high SLNW density is required for realistic
applications. Here, we systematically study the thermoelectric
properties of SLNWs connected to electrodes in the linear and
nonlinear response regimes.

\begin{figure}[h]
\centering
\includegraphics[trim=2.5cm 0cm 2.5cm 0cm,clip,angle=-90,scale=0.3]{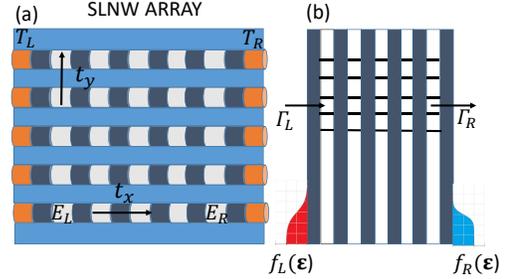}
\caption{(a) Schematic diagram of a quantum dot superlattice
nanowire (SLNW) array connected to electrodes with different
equilibrium temperatures $T_L$ and $T_R$. $t_{x}$ and $t_y$
denote, respectively, the electron hopping strengths in the x and
y directions. For simplicity, only the nearest neighbor hopping
procedure is considered. (b) Band diagram of an SLNW array.
$\Gamma_{L}$ and $\Gamma_R$ denote,respectively, the tunneling
rates of electrons to tunnel from the left and right electrodes
into the outer quantum dots of an SLNW array.}
\end{figure}

\section{Formalism}

To model the thermoelectric properties of an SLNW array, the
Hamiltonian of the system shown in Fig. 1 is given by
$H=H_0+H_{QD}$,[\onlinecite{Haug}] where
\begin{small}
\begin{eqnarray}
H_0& = &\sum_{k,\sigma} \epsilon_k
a^{\dagger}_{k,\sigma}a_{k,\sigma}+ \sum_{k,\sigma} \epsilon_k
b^{\dagger}_{k,\sigma}b_{k,\sigma}\\ \nonumber
&+&\sum_{\ell}^{N_y}\sum_{k,\sigma}
V^L_{k,\ell,j}d^{\dagger}_{\ell,j,\sigma}a_{k,\sigma}
+\sum_{\ell}^{N_y}\sum_{k,\sigma}V^R_{k,\ell,j}d^{\dagger}_{\ell,j,\sigma}b_{k,\sigma}+H.c.
\end{eqnarray}
\end{small}
The first two terms of Eq.~(1) describe the free electron gas in
the left and right electrodes. $a^{\dagger}_{k,\sigma}$
($b^{\dagger}_{k,\sigma}$) creates  an electron of momentum $k$
and spin $\sigma$ with energy $\epsilon_k$ in the left (right)
electrode. $V^L_{k,\ell,j}$ ($V^R_{k,\ell,j}$) describes the
coupling between the left (right) lead with its adjacent QD in the
$\ell$th row, which counts from 1 to $N_y$.
\begin{small}
\begin{eqnarray}
H_{QD}&= &\sum_{\ell,j,\sigma} E_{\ell,j}
d^{\dagger}_{\ell,j,\sigma}d_{\ell,j,\sigma}\\ \nonumber&+&
\sum_{\sigma}\sum_{\ell 1,\ell 2}^{N_y}\sum_{j1,j2}^{N_x} t_{\ell
1,\ell 2, j1, j2} d^{\dagger}_{\ell 1,j1,\sigma} d_{\ell
2,j2,\sigma}+H.c,
\end{eqnarray}
\end{small}
\begin{equation}
t_{\ell 1,\ell 2,j1, j2}= \{ \begin{array}{ll} t_{y,\ell,\ell+1} &
if~j1=j2,  |\ell 1-\ell 2|=1\\
t_{x,j,j+1} & if~\ell 1=\ell 2, |j1-j2|=1
\end{array}.
\end{equation}
where { $E_{\ell,j}$} is the energy level of QD  in the
${\ell}$-th row and $j$-th column. The spin-independent $t_{\ell
1, \ell 2, j1, j2}$ describes the electron hopping strength, which
is limited to the nearest neighboring sites. $d^{\dagger}_{\ell
1,j1,\sigma} (d_{\ell 2,j2,\sigma})$ creates (destroys) one
electron in the QD at the $\ell$th row and $j$th column. If the
wavefunctions of electrons in each QD are localized, the interdot
and intradot Coulomb interactions between electrons are strong.
Their effects on electron transport are significant in the
scenario of weak hopping strengths.[\onlinecite{Kuo5}] On the
other hand, the wave functions of electrons are delocalized in the
scenario of strong hopping strengths, therefore their weak
electron Coulomb interactions can be ignored.[\onlinecite{Lin}]

To study the transport properties of an SLNW array junction
connected to electrodes, it is convenient to use the
Green-function technique. Using the Keldysh-Green's function
technique[\onlinecite{Haug},\onlinecite{Meir}], electron and heat
currents leaving electrodes can be expressed as
\begin{eqnarray}
J &=&\frac{2e}{h}\int {d\epsilon}~
T_{LR}(\epsilon)[f_L(\epsilon)-f_R(\epsilon)],
\end{eqnarray}
and
\begin{eqnarray}
& &Q_{e,L(R)}\\ &=&\frac{\pm 2}{h}\int {d\epsilon}~
T_{LR}(\epsilon)(\epsilon-\mu_{L(R)})[f_L(\epsilon)-f_R(\epsilon)]\nonumber
\end{eqnarray}
where
$f_{\alpha}(\epsilon)=1/\{\exp[(\epsilon-\mu_{\alpha})/k_BT_{\alpha}]+1\}$
denotes the Fermi distribution function for the $\alpha$-th
electrode, where $\mu_\alpha$  and $T_{\alpha}$ are the chemical
potential and the temperature of the $\alpha$ electrode. $e$, $h$,
and $k_B$ denote the electron charge, the Planck's constant, and
the Boltzmann constant, respectively. $T_{LR}(\epsilon)$ denotes
the transmission coefficient of an SLNW array connected to
electrodes, which can be solved by the formula $
T_{LR}(\epsilon)=4Tr[\hat{\Gamma}_{L}\hat{G}^{r}_{D,A}(\epsilon)\hat{\Gamma}_{R}\hat{G}^{a}_{D,A}(\epsilon)]$,
where the matrix of tunneling rates ($\hat{\Gamma}_L$ and
$\hat{\Gamma}_R$) and Green's functions
($\hat{G}^{r}_{D,A}(\epsilon)$ and $\hat{G}^{a}_{D,A}(\epsilon)$)
can be constructed as shown by the example in the
appendix.[\onlinecite{YangNX}]

{In the linear response regime, the electrical conductance ($G_e$)
and Seebeck coefficient ($S$) can be evaluated by using Eq. (4)
with small applied bias $\Delta V=(\mu_L-\mu_R)/e$ and $\Delta
T=T_L-T_R$. We  obtain  $G_e=e^2{\cal L}_{0}$ and $S=-{\cal
L}_{1}/(eT{\cal L}_{0})$. ${\cal L}_n$ is given by
\begin{equation}
{\cal L}_n=\frac{2}{h}\int d\epsilon~
T_{LR}(\epsilon)(\epsilon-E_F)^n\frac{\partial
f(\epsilon)}{\partial E_F},
\end{equation}
where $f(\epsilon)=1/(exp^{(\epsilon-E_F)/k_BT}+1)$ is the Fermi
distribution function of electrodes at equilibrium temperature
$T$.

\section{ Results and discussion}
\subsection{A Single SLNW}
Our discussion begins with a single short SLNW which can be
implemented with current semiconductor fabrication techniques
[\onlinecite{Guerfi}]. In Fig. 2 we calculate the transmission
coefficient $T_{LR}(E_F)$ as a function of QD energy level
($\Delta=E_0-E_F$) with the homogenous electron hopping strength
of $t_{x,j,j+1}=t_c=6\Gamma_0$  for an SLNW with QD number
$N_x=N=5$ and one energy level for each QD. All energy scales are
in units of $\Gamma_0=1~meV$ through out this article. Meanwhile,
we have adopted symmetrical tunneling rates
$\Gamma_L=\Gamma_R=\Gamma$. Diagrams (a),(b) and (c) consider
different tunneling rates of $\Gamma=1\Gamma_0, 3\Gamma_0$ and
$6\Gamma_0$, respectively. Fig. 2(a) clearly shows the electronic
structures of a single SLNW. The resonant channels of Fig. 2(a)
are given by $\epsilon=E_0-2t_c~ cos(\frac{n\pi}{N+1})$ with
$n=1,2,..N$, which is a simple tight-binding outcome with
non-periodic boundary condition and ignores the effect of
electrodes. Electron transport in Fig. 2(a) illustrates QD Fabry
Perot type oscillations.[\onlinecite{Whitney2}] When $\Gamma$
increases up to $6\Gamma_0$, the electronic structure of SLNW can
not be resolved completely. We note that the resonant channels
predicted by $\epsilon=E_0-2t_c~ cos(\frac{n\pi}{N+1})$ are
shifted in Fig. 2(c) due to the strong coupling between the outer
QDs and the electrodes. The electronic structure of N-QDs shows
$N-2$ resonant channels at large tunneling rates
($\Gamma=12\Gamma_0=2t_c$). Such a behavior results from that the
outer QDs replace the role of electrodes when $\Gamma \ge 2t_c$.
The results of Fig. 2 indicate that the distribution of
$T_{LR}(\epsilon)$ depends on $t_{x,j,j+1}$ and $\Gamma$ values.
Ref. [\onlinecite{Whitney1}] pointed out that the maximum
efficiency of heat engines with finite output power will be
reached when the transmission coefficient maintains a square form
. However, it is not yet clear how a $ T_{LR}(\epsilon)$ with
square form may be constructed. A single quantum dot chain has
been proposed to realize the boxcar form of $T_{LR}(\epsilon)$
[\onlinecite{Whitney2}], but its $G_e$, $S$ and power factor
$PF=S^2G_e$ are lacking.
\begin{figure}[h]
\centering
\includegraphics[angle=-90,scale=0.3]{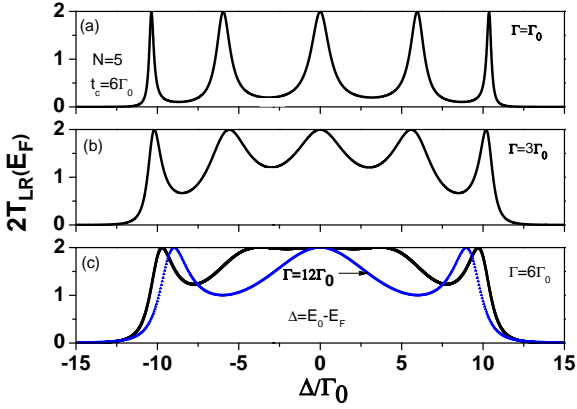}
\caption{Transmission coefficient $T_{LR}(E_F)$ as a function of
QD energy level for $N=5$ and $t_{x,j,j+1}=t_c=6\Gamma_0$.
Diagrams (a), (b) and (c) are for
$\Gamma=1\Gamma_0$,$\Gamma=3\Gamma_0$ and $\Gamma=6\Gamma_0$ in
that order. The extra curve in diagram (c) (triangles) for
$\Gamma=12\Gamma_0$.}
\end{figure}

To examine the effect of boxcar form of $T_{LR}(\epsilon)$ on the
thermoelectric properties of
SLNWs,[\onlinecite{Whitney1},\onlinecite{Whitney2}] we calculate
$G_e$, $S$ and $PF$  by considering inhomogenous electron hopping
strengths in Fig. 3. The electron hopping strengths
$t_{12}=t_{45}=0.78\Gamma$ and $t_{23}=t_{34}=0.56\Gamma$ are
adopted for $N=5$ and $\Gamma=6\Gamma_0$. The curve with triangle
marks of $G_e$ (at $k_BT=0$) in Fig. 3(a) corresponds to the
boxcar form transmission coefficient.[\onlinecite{Whitney2}] The
temperature-dependent $G_e$ shows a typical thermal broadening
feature. The Seebeck coefficient is extremely small in the highly
conductive region ($|\Delta| \le 6 \Gamma_0$). The Seebeck
coefficients have different signs for positive and negative
$\Delta$ values. The negative (positive) $S$ indicates that the
electron transports of electrodes are mainly dominated by the
resonant channels above (below) the Fermi energy of electrodes. In
general, electrons of electrodes tunneling through the resonant
channels below $E_F$ are called holes. Therefore, the change of
sign in the Seebeck coefficients is called bipolar
behavior.[\onlinecite{Kuo5}] The peak position of $PF$ shifts away
from $E_F$ when the temperature increases.

\begin{figure}[h]
\centering
\includegraphics[angle=-90,scale=0.3]{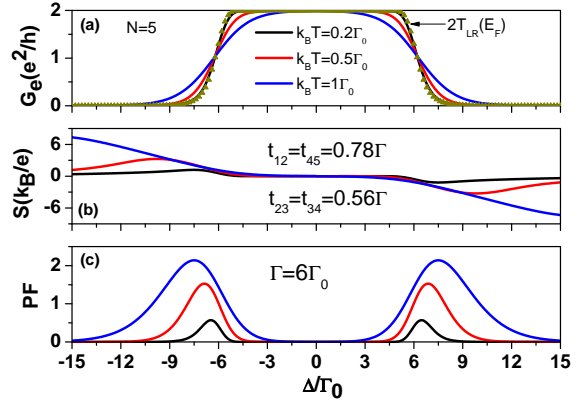}
\caption{(a) Electrical conductance, (b) Seebeck coefficient and
(c) power factor $PF$ as a function of $\Delta=E_0-E_F$ for
different temperature at $N=5$, $\Gamma=6\Gamma_0$,
$t_{12}=t_{45}=0.78\Gamma$ and $t_{23}=t_{34}=0.56\Gamma$ . The
$PF$ is in the units of $k^2_B/h$. The curve with triangle marks
in Fig. 3(a) corresponds to the boxcar-form transmission
coefficient ($G_e=\frac{2e^2}{h}T_{LR}(E_F)$).}
\end{figure}

Now we discuss in detail the differences of power factor between
the transmission coefficients in Fig. 2 and Fig. 3.  In Fig. 4
$G_e$, $S$ and $PF$ as a function of temperature for different
$\Delta $ values at $N=5$ are calculated. Diagrams (a),(b) and (c)
consider the quasi-square form $T_{LR}(\epsilon)$ given at the
condition of $t_{x,j,j+1}=t_c$ and $t_c=\Gamma=6\Gamma_0$. In the
quasi-square form, $G_e$ drops quickly for $\Delta=10\Gamma_0$
when $k_BT$ is below 1.5$\Gamma_0$. Such a behavior is attributed
to the electron transport mainly resulting from resonant tunneling
procedure and the electron population below $E_F$ is reduced with
increased temperature. When $\Delta=20\Gamma_0$ and
$\Delta=30\Gamma_0$ (resonant channels are far away from the $E_F$
of electrodes), the electron transports between the electrodes are
dominated by the thermionically-assisted tunneling procedure
(TATP).[\onlinecite{Mahan}] For diagrams (d),(e) and (f), the
curves of $G_e$, $S$ and $PF$ correspond to the boxcar
transmission function in Fig. 3. If the curve of Fig. 4(c) at
$\Delta=10\Gamma_0$ is compared with that of Fig. 4(f), the
maximum $PF$ given by the boxcar form is larger than that of the
quasi-square form. For $\Delta=10\Gamma_0$, the Seebeck
coefficient of the boxcar form is much larger than that of the
quasi-square form. For two other cases $\Delta=20\Gamma_0$ and
$\Delta=30\Gamma_0$, the $PF$ of the quasi-square form is better
than that of the boxcar form. As for the electron Coulomb
interactions, which are important for SLNWs in the Coulomb
blockade regime, we have demonstrated that $PF$ is reduced in the
presence of Coulomb interactions.[\onlinecite{Kuo5}]

\begin{figure}[h]
\centering
\includegraphics[angle=-90,scale=0.3]{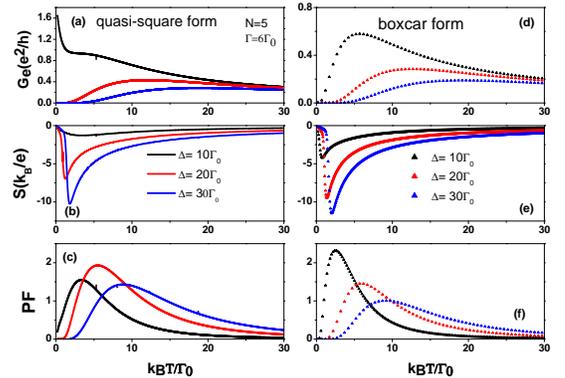}
\caption{(a) Electrical conductance, (b) Seebeck coefficient and
(c) power factor as a function of $k_BT$ for different $\Delta$
values at $N=5$, $t_{j,j+1}=6\Gamma_0$ and $\Gamma=6\Gamma_0$. The
curves of diagrams (d), (e) and (f) consider the boxcar-form
transmission coefficient in Fig. 3.}
\end{figure}

\subsection{An SLNW Array}
Although many studies have investigated the phonon thermal
conductivity of 2-dimensional
systems,[\onlinecite{ChenG}-\onlinecite{GuoYY}] the thermoelectric
properties of an SLNW array are lacking. Fig. 5 shows the
transmission coefficient $T_{LR}(E_F)$ as a function of dot energy
level $E_{\ell,j}=E_0=\Delta+E_F$ at $N_x=N_y=5$, where $N_x$ and
$N_y$ are quantum dot numbers in the x and y directions,
respectively. For $t_{y,\ell,\ell+1}=t_y=0$ and
$t_{x,j,j+1}=t_x=6\Gamma_0$, $T_{LR}(E_F)$ shows the maximum
probability for the electron transport between the electrodes. The
ranges of $T_{LR}(E_F)$ are highly enhanced with increasing $t_y$.
Fig. 5(a) illustrates the transition between a one dimensional
system and a two dimensional system. The feature of
$T_{LR}(\epsilon)$ involves the electronic structure of SLNW
arrays given by $\epsilon=E_0-2t_x~cos(\frac{n_x\pi}{N_x+1})-2t_y~
cos(\frac{n_y\pi}{N_y+1})$, where $n_x=1,2,..N_x$ and
$n_y=1,2,..N_y$. If QDs have stronger coupling strengths in the y
direction ($t_y > t_x$), how such a geometry is to influence the
transport behavior of electrons. To further reveal the situation
of $t_y > t_x$, we plot $T_{LR}(E_F)$ in Fig. 5 (b) with
$t_y=12\Gamma_0$ and $t_x=1\Gamma_0$. Each substructure of a main
structure exhibits features similar to the structure of $t_y=0$
and $t_x=6\Gamma_0$ in Fig. 5(a). The behavior of Fig. 5(b) can be
regarded as a single SLNW with multiple energy levels in each QD.

\begin{figure}[h]
\centering
\includegraphics[angle=-90,scale=0.3]{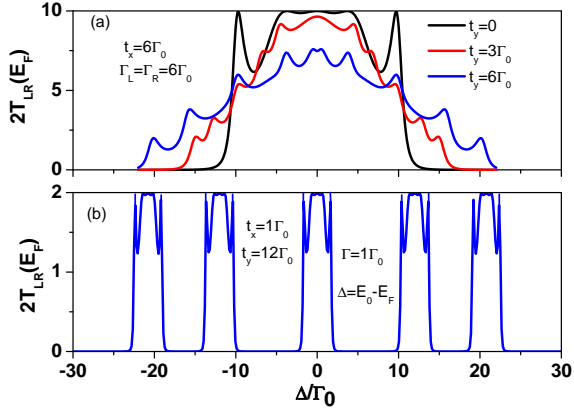}
\caption{Transmission coefficient as a function of quantum dot
energy level ($E_{\ell,j}=E_0-E_F=\Delta$) for $N_x=N_y=5$
.Diagram (a) considers $t_{x,j,j+1}=t_x=6\Gamma_0$ for different
$t_{y,\ell,\ell+1}=t_y$ values at $\Gamma_L=\Gamma_R=6\Gamma_0$,
and diagram (b) considers $t_x=1\Gamma_0$, $t_y=12\Gamma_0$ and
$\Gamma_L=\Gamma_R=1\Gamma_0$.}
\end{figure}

To examine the effects of $t_y$ on the thermoelectric properties
of an SLNW array, we calculate $G_e$, $S$ and $PF$ as a function
of QD energy level $\Delta=E_0-E_F$ for different $t_y$ values at
low temperature $k_BT=1\Gamma_0$ in Fig. 6. The behavior of $G_e$
at low temperature is significantly different from that at zero
temperature ($G_e=\frac{2e^2}{h}T_{LR}(E_F)$), however resonant
tunneling procedure (RTP) still dominates the electron transport
between the electrodes. $S$ is vanishingly small in highly
conductive region whether the SLNW array is in the 1-D or 2-D
topological structures. In addition, the maximum $S$ value of
$t_y=0$ is the same as that of $t_y=6\Gamma_0$. In Fig. 6(c) the
maximum $PF$ value is given by $PF_1$ for $t_y=0$. The results of
Fig. 6 indicate that the $PF$ of the 1-D system ($t_y=0$) is
better than that of 2-D system ($t_y=6\Gamma_0$) when the RTP
dominates the electron transport.

\begin{figure}[h]
\centering
\includegraphics[angle=-90,scale=0.3]{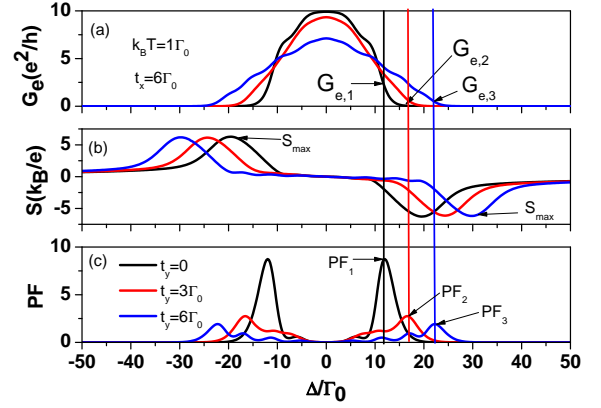}
\caption{(a) Electrical conductance, (b) Seebeck coefficient and
(c) power factor as a function of $\Delta$ for different $t_y$
values at $k_BT=1\Gamma_0$. Other physical parameters are the same
as those of Fig. 5(a).}
\end{figure}

Because many thermoelectric devices operate at high
temperatures,[\onlinecite{Chen}] we examine the effects of
electron hopping strengths between SLNWs on $PF$ in a large
temperature range. Fig. 7 shows $S$ and $PF$ as functions of
temperature for two different $t_y$ values. Diagrams (a) and (b)
consider the case of $\Delta=10\Gamma_0$. Diagrams (c) and (d)
consider $\Delta=30\Gamma_0$. The behaviors of $G_e$ at $t_y=0$
can be referred to the curves of Fig. 4(a). The trend of maximum
$PF$ with respect to $t_y$ is the same that of $S$, because $G_e$
is not sensitive to $t_y$ at $\Delta=10\Gamma_0$ when $k_BT \ge
2.5\Gamma_0$.  For $\Delta=10\Gamma_0$, we have the ratio of
$PF_{1-D}/PF_{2-D}=3.9$. For $\Delta=30\Gamma_0$,
$PF_{1-D}/PF_{2-D}$ is near one. As $\Delta$ is increased up to
$\Delta=60\Gamma_0$, the topological effect nearly vanishes
($PF_{1-D}/PF_{2-D}=1$). This implies that the optimization of
$PF$ in a 1-D system is still useful for a 2-D system as long as
$\frac{t_y}{\Delta}\le 0.1$.

\begin{figure}[h]
\centering
\includegraphics[angle=-90,scale=0.3]{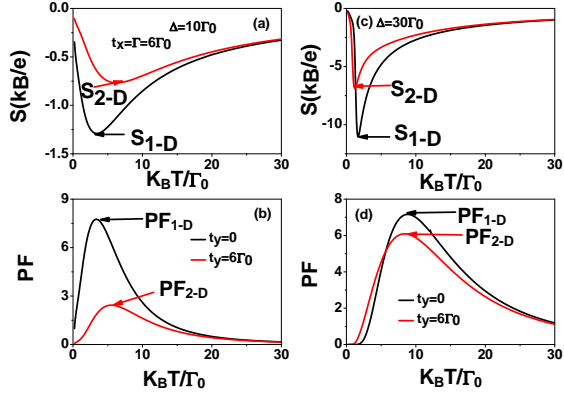}
\caption{(a) Seebeck coefficient and (b) power factor as a
function of temperature for different $t_y$ values at
$\Delta=10~\Gamma_0$. (c) Seebeck coefficient and (d) power factor
as a function of temperature for different $t_y$ values at
$\Delta=30~\Gamma_0$. $t_x=6~\Gamma_0$ and
$\Gamma_L=\Gamma_R=6~\Gamma_0$.}
\end{figure}

\subsection{Electron Heat Rectification}
Recently, many theoretical studies have devoted to the design of
heat diodes (HDs).[\onlinecite{Terraneo}-\onlinecite{Craven}]
Those designs employed three kind of heat carriers, including
phonons,[\onlinecite{Terraneo}-\onlinecite{Segal}]
photons[\onlinecite{Otey}] and
electrons[\onlinecite{Kuo1},\onlinecite{Craven}]. So far, most
experimental findings of heat rectification ratios fall between 1
and 1.4.[\onlinecite{ChangCW}]Although high rectification ratio
for electron HD was reported in metal/superconductor junction
systems operating at extremely low temperatures (below
liquid-helium temperature),[\onlinecite{Perez}] it is desirable to
investigate whether the SLNW arrays can show such a functionality.
In Eq. (5), $Q_{e,L}+Q_{e,R}=-(\mu_L-\mu_R)\times J/e$, which
describes the Joule heating. To discuss the electron heat
rectification, we consider the open circuit condition ($J=0$)
under a temperature bias $\Delta T=T_L-T_R$, where $T_L=T+\Delta
T/2$ and $T_R=T-\Delta T/2$. For $J=0$, $Q_{e,L}(\Delta
T)=-Q_{e,R}(\Delta T)=Q_e(\Delta T)$, in which the contribution
involving $\mu_{L(R)}$ is zero. Due to the Seebeck effect, the
thermal voltage $V_{th}$ induced by $\Delta T$ will balance the
electrons diffused from the hot electrode to the cold electrode to
establish the condition of
$J=0$.[\onlinecite{Kuo1},\onlinecite{Craven}] Meanwhile, the
energy levels $E_{\ell,j}$ will be modified due to the presence of
$V_{th}$.[\onlinecite{Craven}] Consequently, $T_{LR}(\epsilon)$
will depend on $V_{th}$. In addition, the Fermi distribution
functions ($f_{L(R)}(\epsilon)$) also depend on the thermal
voltage ($\mu_L=E_F+eV_{th}/2$ and $\mu_R=E_F-eV_{th}/2$).

\begin{figure}[h]
\centering
\includegraphics[angle=-90,scale=0.3]{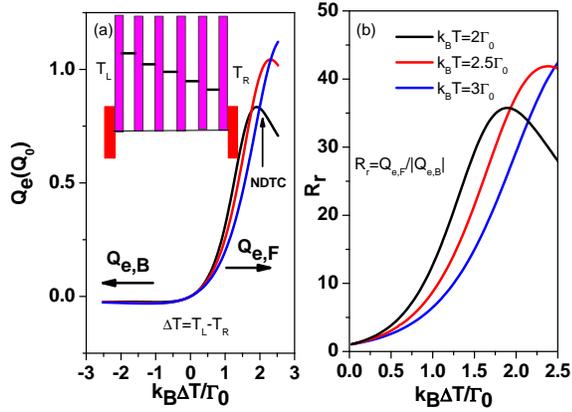}
\caption{(a) Electron heat current and (b) heat rectification
ratio as a function of temperature bias for various values of $T$
at $N_x=N_y=5$, $t_x=1~\Gamma_0$,  $\Gamma=1~\Gamma_0$, $t_y=0$,
$\Delta E=0.8~\Gamma_0$ and $E_{\ell,R}=E_F+4\Gamma_0$.
$Q_0=\Gamma^2_0/h$. }
\end{figure}

We consider each SLNW with a staircase alignment of energy levels
(see the inset of Fig. 8). Each QD has the position-dependent
energy level only in the x-direction:
$E_{\ell,j}=E_R+(N_x-j)\Delta E$, where $\Delta E$ denotes the
energy level separation. Such a variation in QD levels can be
engineered by considering suitable size variation of QDs in the
SLNW.[\onlinecite{Guerfi}] We consider an SLNW array with $N_x=5$
and $N_y=5$. Namely, we have $E_{\ell,1}=E_R+4\Delta E$,
$E_{\ell,2}=E_R+3\Delta E$... and $E_{\ell,5}=E_R$. With an
induced thermal voltage, $V_{th}$, the energy levels $E_{\ell,j}$
are modified as $\varepsilon_{\ell,j}=E_{\ell,j}+\eta_D eV_{th}$.
In a simple approximation where the electric field is uniformly
distributed in the x-direction, the level modulation factor is
expressed as $\eta_{D}=-(j-3)*Ls/L$. The pair length (that of one
QD plus one spacer layer) is $L_s$ and the length of a single SLNW
is $L$. We have used their ratio $L_s/L=0.2$[\onlinecite{Hu}]. The
thermal voltage ($V_{th}$) can be evaluated by Eq.~(4) under the
condition of $J=0$. Once $V_{th}$ is obtained, the electron heat
currents $Q_e(\Delta T)$ can be evaluated by Eq.~(5). The
resulting $Q_e$ as a function of temperature bias for various
values of $T$ at $t_x=1~\Gamma_0$, $\Gamma=1\Gamma_0$ and $t_y=0$
is plotted in Fig.~8(a). In Fig. 8(a) $Q_e$ shows the features of
thermal conductors and thermal insulators under the forward
temperature bias ($\Delta T > 0$) and reverse temperature bias
($\Delta T < 0$), respectively. The electron heat rectification
ratio of $R_r=\frac{Q_{e}(\Delta T>0)}{|Q_{e}(\Delta
T<0)|}=\frac{Q_{e,F}}{|Q_{e,B}|}$ is plotted in Fig. 8(b). Because
$Q_{e,B}$ is insensitive to the variation of $\Delta T$, the
behavior of $R_r$ is very similar to $Q_{e,F}$. To design HDs, one
needs to have a high $R_r$ at a small temperature bias
[\onlinecite{Chiu}]. Fig. 8(b) shows that $R_r$ is larger than ten
at a small temperature bias ($\Delta T/T=0.5$).

The asymmetric behavior of $Q_e$ can be understood by considering
$V_{th}$ as a function of $\Delta T$. $Q_e$ and $V_{th}$ as
functions of temperature bias for different $\Delta E$ values at
$k_BT=2\Gamma_0$ are plotted in Fig. 9. One see that the
asymmetrical behavior of $Q_e$ only exists for $\Delta E \ne 0$.
This implies that the staircase energy levels of SLNWs play a
remarkable role in observing the electron heat rectification.
Using the curve of $\Delta E=0.8\Gamma_0$ to illustrate the heat
rectification, the QD levels are nearly aligned at $k_B\Delta T
=1.75\Gamma_0$, which gives $eV_{th}=-4~\Gamma_0$, allowing the
resonant tunneling of electrons from the left electrode to the
right electrode. When $k_B\Delta T
> 1.75\Gamma_0$, $\varepsilon_L=E_R+4\Delta E+0.4eV_{th}$ and $\varepsilon_R=E_R-0.4eV_{th}$ are
off-resonant, which explains why the negative differential thermal
conductance (NDTC) occurs at $k_B\Delta T > 1.75\Gamma_0$.
Meanwhile  the QD levels are misaligned under a reverse
temperature bias, leading to an off-resonance condition (see
insets in Fig. 9(b)).

\begin{figure}[h]
\centering
\includegraphics[angle=-90,scale=0.3]{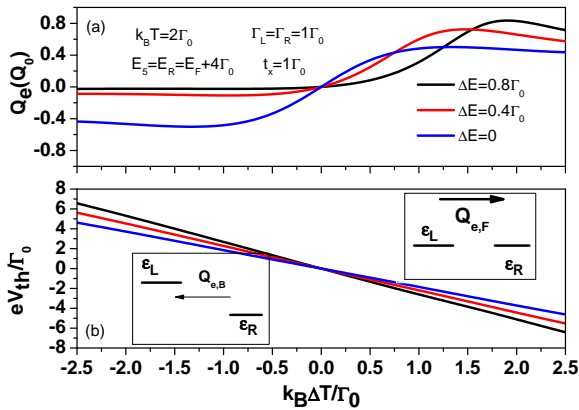}
\caption{ (a) Electron heat current and (b) thermal voltage as a
function of temperature bias for various values of $\Delta E$ at
$K_BT=2\Gamma_0$. Other physical parameters are the same as those
of Fig. 8.}
\end{figure}

In Figures (8) and (9), we have considered the case of $t_y=0$. To
clarify the effects of $t_y$ on electron heat diodes, we calculate
the electron heat current and heat rectification ratio as a
function of temperature bias for different $t_y$ values at
$k_BT=3~\Gamma_0$ in Fig. 10. $Q_e$ increases with increasing
$t_y$ values. For a finite $t_y$ value, the degeneracy between
$E_{\ell,j}$ in y-direction is destroyed. The rectification
behavior of the SLNW array can be regarded as that of a single
SLNW with "multi energy levels in each QD". Different energy
levels provide the electrons with different kinetic energies. With
increasing temperature bias ($\Delta T > 0$), these multi-energy
levels in each QD form the multi-subbands, which substantially
enhance the electron heat currents. Because of a small $\Delta E$,
these multi energy levels resulting from a finite $t_y$ still
provide some paths for electron transport in the reverse
temperature bias to increase $Q_e$. Although the $R_r$ values are
much reduced with increasing $t_y$ in Fig. 10(b), they are still
very impressive when compared to some experimental
findings.[\onlinecite{ChangCW}]

\begin{figure}[h]
\centering
\includegraphics[angle=-90,scale=0.3]{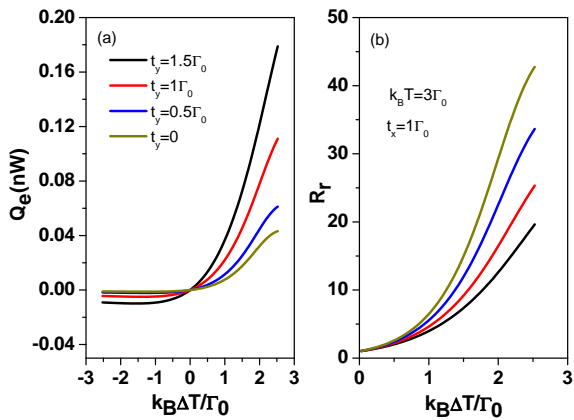}
\caption{(a)Electron heat current and (b) heat rectification
ratio, $R_r$ as a function of temperature bias $\Delta T$ for
various values of $t_y$ at $k_BT=3\Gamma_0$ and $\Delta
E=0.8~\Gamma_0$. Other physical parameters are the same as those
of Fig. 9.Note that we use $nW$ to describe the magnitude of $Q_e$
instead of $Q_0$ for $\Gamma_0=1~meV$.}
\end{figure}
Next, we will demonstrate that the TATP also plays an important
role in observing electron heat rectification. Figures 11(a) and
11(b), show $R_r$ and $Q_e$ as a function of temperature bias for
different $E_R$ values, respectively. For $E_R=E_F$, the maximum
$R_r$ is smaller than three. When $E_R$ is far away from $E_F$,
the maximum $R_r$ reaches 30. Nevertheless, the magnitude of $Q_e$
is severely suppressed for large $E_R$ values. The behavior of
NDTC can be observed for the cases of $\Delta_R=8~\Gamma_0$ and
$\Delta_R=12~\Gamma_0$. To further clarify the results of diagram
(b), we show $Q_e$ and the Seebeck coefficient ($S=V_{th}/\Delta
T$) as functions of $E_R$ for different $k_BT$ values at
$k_B\Delta T=2~\Gamma_0$ in diagrams (c) and (d). $Q_e$ decays
quickly with increasing $\Delta_R=E_R-E_F$, whereas $V_{th}=\Delta
T~S$ is much enhanced. When $\Delta_R > 5\Gamma_0$, the TATP
dominates electron transport due to all resonant channels being
above $E_F$. The curves of $\Delta_R=8\Gamma_0$ and
$\Delta_R=12\Gamma_0$ in Fig. 11(a) demonstrate that the TATP as
well as $\Delta E \neq 0$ plays a critical role in observing heat
rectification.

\begin{figure}[h]
\centering
\includegraphics[angle=-90,scale=0.3]{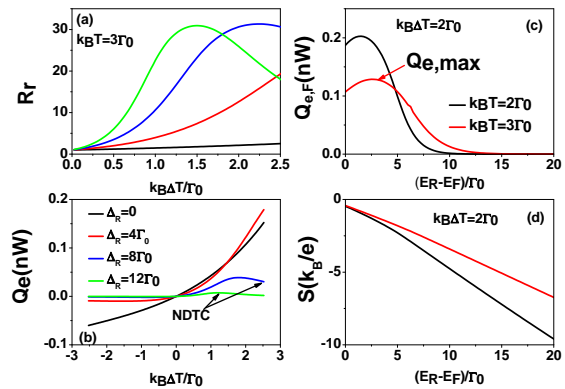}
\caption{(a) Heat rectification ratio and (b) electron heat
current as a function of temperature bias for different
$\Delta_R=E_R-E_F$ values at $t_x=1\Gamma_0$, $t_y=1.5~\Gamma_0$,
$\Delta E=0.8~\Gamma_0$ and $\Gamma=1~\Gamma_0$. (c) Electron heat
current and (d) nonlinear Seebeck coefficient ($S=V_{th}/\Delta
T$) as a function of $E_R$ for two different average temperatures
at $k_B\Delta T=2~\Gamma_0$. }
\end{figure}

\section{Conclusion}
The thermoelectric properties of an SLNW array connected to
metallic electrodes are theoretically studied by using the
tight-binding Hamiltonian combined with the nonequilibrium Green's
function method. The electron current and heat current are
significantly influenced by their transmission coefficients, which
depend on the electron hopping strengths, QD energy levels and
electron tunneling rates. These physical parameters determined by
the shape and size of each QD can be calculated in the framework
of effective mass theory for semiconductor QDs.[\onlinecite{Kuo8}]
The effects of electron interwire hopping on the optimization of
power factor can be ignored if the TATP dominates the electron
transport between the electrodes. In the nonlinear regime electron
heat current can be highly enhanced due to proximity effect,
whereas the electron heat rectification ratio is suppressed.
Finally, we find that the TATP as well as staircase energy levels
distributed in QDs play a very important role in observing the
electron HDs with high $R_r$ values.
%\begin{flushleft}

%\end{flushleft}

%\begin{flushleft}
{\bf Acknowledgments}\\
%\end{flushleft}
This work was supported under Contract No. MOST 107-2112-M-008
-023MY2
\mbox{}\\
E-mail address: mtkuo@ee.ncu.edu.tw\\
%E-mail address: yiachang@gate.sinica.edu.tw\\

%\renewcommand{\thesection}{\mbox{Appendix s}} %\section{Appendix}~\Roman{section}
%\setcounter{section}{0}

%\renewcommand{\theequation}{\mbox{A.\arabic{equation}}} %\section{Appendix}
%\setcounter{equation}{0} % reset counter

%\section{}
%\subsection{Derivation of the tunneling current formula using Dyson's equations\label{App:TC_l} }
\mbox{}\\
%{\bf Appendix A. }

\section{Appendix}
When $N_x=N_y=2$, we have $E_{1,1}=E_1$, $E_{1,2}=E_2$,
$E_{2,1}=E_3$ and $E_{2,2}=E_4$. The electron hopping strength
between $E_{1(2)}$ ($E_{1(3)}$) and $E_{3(4)}$ ($E_{2(4)}$) is
denoted by $t_y$ ($t_x$). The transmission coefficient
$T_{LR}(\epsilon)$ of an SLNW array is calculated by the formula $
T_{LR}(\epsilon)=4Tr[\hat{\Gamma}_{L}\hat{G}^{r}_{D,A}(\epsilon)\hat{\Gamma}_{R}\hat{G}^{a}_{D,A}(\epsilon)]$,[\onlinecite{Haug},\onlinecite{YangNX}]
where tunneling rates $\hat{\Gamma}_L$ and $\hat{\Gamma}_R$ are
assumed to be energy-independent. Their forms are given by
\begin{small}
\begin{equation}
\hat{\Gamma}_L=\Gamma_L \left[
\begin{array}{cccc}
1&0&0&0\\
0&0&0&0\\
0&0&1&0\\
0&0&0&0
\end{array}\right],
\end{equation}
\end{small}
and
\begin{small}
\begin{equation}
\hat{\Gamma}_R=\Gamma_R \left[
\begin{array}{cccc}
0&0&0&0\\
0&1&0&0\\
0&0&0&0\\
0&0&0&1
\end{array}\right].
\end{equation}
\end{small}
From Eqs. (A.1) and (A.2), $E_1$ and $E_3$ ($E_2$ and $E_4$) are
coupled to the left (right) electrode.
$\Gamma_{L(R)}=\pi\sum_{k}|V^{L(R)}_{k,\ell,j}|^2\delta(\epsilon-\epsilon_k)$.
$\hat{G}^{r}_{D,A}(\epsilon)$ and $\hat{G}^{a}_{D,A}(\epsilon)$
can be calculated by their inverse matrixes
($\hat{G}^{r^{-1}}_{D,A}(\epsilon)$ and
$\hat{G}^{a^{-1}}_{D,A}(\epsilon)$), which are

\begin{small}
\begin{eqnarray}
&&\hat{G}^{r^{-1}}_{D,A}(\epsilon)\\
&=&\left[
\begin{array}{cccc}
\epsilon-E_1+i\Gamma_L&t_{x}&t_{y}&0\\
t_{x}&\epsilon-E_2+i\Gamma_R&0&t_{y}\\
t_{y}&0&\epsilon-E_3+i\Gamma_L&t_{x}\\
0&t_{y}&t_{x}&\epsilon-E_4+i\Gamma_R
\end{array}\right]\nonumber,
\end{eqnarray}
\end{small}
and

\begin{small}
\begin{eqnarray}
&&\hat{G}^{a^{-1}}_{D,A}(\epsilon)\\
&=&\left[
\begin{array}{cccc}
\epsilon-E_1-i\Gamma_L&t_{x}&t_{y}&0\\
t_{x}&\epsilon-E_2-i\Gamma_R&0&t_{y}\\
t_{y}&0&\epsilon-E_3-i\Gamma_L&t_{x}\\
0&t_{y}&t_{x}&\epsilon-E_4-i\Gamma_R
\end{array}\right]\nonumber.
\end{eqnarray}
\end{small}
The imaginary parts of diagonal matrix elements result from the
coupling between QDs and electrodes. Off-diagonal matrix elements
($t_x$ and $t_y$) present the electron hopping strengths between
QDs. Only the nearest neighbor's hopping strengths are included in
Eqs. (A.3) and (A.4). After tedious algebra,
$\hat{G}^{r}_{D,A}(\epsilon)$ is written as
\begin{small}
\begin{eqnarray}
&&\hat{G}^{r}_{D,A}(\epsilon)\\
&=&\frac{1}{D}\left[
\begin{array}{cccc}
a_{11}&a_{12}&a_{13}&a_{14}\\
a_{21}&a_{22}&a_{23}&a_{24}\\
a_{31}&a_{32}&a_{33}&a_{34}\\
a_{41}&a_{42}&a_{43}&a_{44}
\end{array}\right]\nonumber,
\end{eqnarray}
\end{small}
where we have
\begin{small}
\begin{eqnarray}
& &D\\ \nonumber
&=&(\epsilon-E_1+i\Gamma_L)(\epsilon-E_2+i\Gamma_R)(\epsilon-E_3+i\Gamma_L)(\epsilon-E_4+i\Gamma_R)\\
\nonumber
&-&t^2_y(\epsilon-E_1+i\Gamma_L)(\epsilon-E_3+i\Gamma_L)\\
\nonumber
&-&t^2_y(\epsilon-E_2+i\Gamma_R)(\epsilon-E_4+i\Gamma_R)\\
\nonumber
&-&t^2_x(\epsilon-E_1+i\Gamma_L)(\epsilon-E_2+i\Gamma_R)\\
\nonumber
&-&t^2_x(\epsilon-E_3+i\Gamma_L)(\epsilon-E_4+i\Gamma_R)\\
&+&(t^2_x-t^2_y)^2 \nonumber
\end{eqnarray}
\end{small}
and
\begin{small}
\begin{eqnarray}
&&a_{11}\\ \nonumber
&=&(\epsilon-E_2+i\Gamma_R)(\epsilon-E_3+i\Gamma_L)(\epsilon-E_4+i\Gamma_R)\\
\nonumber
&-&t^2_y(\epsilon-E_3+i\Gamma_L)-t^2_x(\epsilon-E_2+i\Gamma_R)\\
\nonumber
a_{12}&=&t^3_x-t^2_yt_x-t_x(\epsilon-E_3+i\Gamma_L)(\epsilon-E_4+i\Gamma_R)\\
\nonumber a_{13}&=&t^3_{y}-t^2_xt_y-t_y(\epsilon-E_2+i\Gamma_R)(\epsilon-E_4+i\Gamma_R)\\
\nonumber
a_{14}&=&t_xt_y(\epsilon-E_2+i\Gamma_R)+t_xt_y(\epsilon-E_3+i\Gamma_L)\\
\nonumber
a_{21}&=&
t^3_x-t^2_yt_x-t_x(\epsilon-E_3+i\Gamma_L)(\epsilon-E_4+i\Gamma_R)\\
\nonumber a_{22}
&=&(\epsilon-E_1+i\Gamma_L)(\epsilon-E_3+i\Gamma_L)(\epsilon-E_4+i\Gamma_R)\\
\nonumber
&-&t^2_x(\epsilon-E_1+i\Gamma_L)-t^2_y(\epsilon-E_4+i\Gamma_R)\\
\nonumber
a_{23}&=&t_yt_x(\epsilon-E_1+i\Gamma_L)+t_{y}t_x(\epsilon-E_4+i\Gamma_R)\\
\nonumber
a_{24}&=&t^3_y-t^2_xt_y-t_x(\epsilon-E_1+i\Gamma_L)(\epsilon-E_3+i\Gamma_L)\\
\nonumber
a_{31}&=&t^3_y-t^2_xt_y-t_y(\epsilon-E_2+i\Gamma_R)(\epsilon-E_4+i\Gamma_R)\\
\nonumber
a_{32}&=&t_yt_x(\epsilon-E_1+i\Gamma_L)+t_yt_x(\epsilon-E_4+i\Gamma_R)\\
\nonumber
a_{33}&=&(\epsilon-E_1+i\Gamma_L)(\epsilon-E_2+i\Gamma_R)(\epsilon-E_4+i\Gamma_R)\\
\nonumber
&-&t^2_y(\epsilon-E_1+i\Gamma_L)-t^2_x(\epsilon-E_4+i\Gamma_R)\\
\nonumber
a_{34}&=&t^3_x-t^2_yt_x-t_x(\epsilon-E_1+i\Gamma_L)(\epsilon-E_2+i\Gamma_R)\\
\nonumber
a_{41}&=&t_xt_y(\epsilon-E_2+i\Gamma_R)+t_xt_y((\epsilon-E_3+i\Gamma_L)\\
\nonumber
a_{42}&=&t^3_y-t^2_xt_y-t_x(\epsilon-E_1+i\Gamma_L)(\epsilon-E_3+i\Gamma_L)\\
\nonumber
a_{43}&=&t^3_x-t^2_yt_x-t_x(\epsilon-E_1+i\Gamma_L)(\epsilon-E_2+i\Gamma_R)\\
\nonumber
a_{44}&=&(\epsilon-E_1+i\Gamma_L)(\epsilon-E_2+i\Gamma_R)(\epsilon-E_3+i\Gamma_L)\\
\nonumber
&-&t^2_y(\epsilon-E_2+i\Gamma_R)-t^2_x(\epsilon-E_3+i\Gamma_L).\\
\nonumber
\end{eqnarray}
\end{small}
Although there are $16$ matrix elements, many off-diagonal matrix
elements are the same. Likewise, we can obtain
$\hat{G}^a_{D,A}(\epsilon)$. Using Eqs. (A.1), (A.2) and (A.5),
the closed form of $T_{RL}(\epsilon)$ is obtained by some
algebraic maneuvers. We have

\begin{equation}
T_{LR}(\epsilon)=\frac{c_{12}+c_{14}+c_{34}+c_{32}}{|D|^2}
\end{equation}
where
\begin{small}
\begin{eqnarray}
& &c_{12}\\ \nonumber
&=&4\Gamma_Lt^2_x\Gamma_R|t^2_x-t^2_y-(\epsilon-E_3+i\Gamma_L)(\epsilon-E_4+i\Gamma_R)|^2\\
\nonumber
c_{14}&=&4\Gamma_Lt^2_xt^2_y\Gamma_R|(\epsilon-E_2+i\Gamma_R)+(\epsilon-E_3+i\Gamma_L)|^2\\
\nonumber
c_{34}&=&4\Gamma_Lt^2_x\Gamma_R|t^2_x-t^2_y-(\epsilon-E_1+i\Gamma_L)(\epsilon-E_2+i\Gamma_R)|^2\\
\nonumber
c_{32}&=&4\Gamma_Lt^2_xt^2_y\Gamma_R|(\epsilon-E_1+i\Gamma_L)+(\epsilon-E_4+i\Gamma_R)|^2\\
\nonumber.
\end{eqnarray}
\end{small}
For $t_y=0$, $T_{LR}(\epsilon)$ is given by the simple expression
below
\begin{small}
\begin{eqnarray}
& &T_{LR}(\epsilon) \\ \nonumber
&=&4(\frac{\Gamma_Lt^2_x\Gamma_R}{|(\epsilon-E_1+i\Gamma_L)(\epsilon-E_2+i\Gamma_R)-t^2_x|^2}\\
\nonumber
&+&\frac{\Gamma_Lt^2_x\Gamma_R}{|(\epsilon-E_3+i\Gamma_L)(\epsilon-E_4+i\Gamma_R)-t^2_x|^2}).
\end{eqnarray}
\end{small}
Eq. (A.10) illustrates the $T_{LR}(\epsilon)$ of two parallel
serially coupled quantum dots in the absence of $t_y$. For
$N_x=N_y=N > 2$, $\hat{\Gamma}_L$, $\hat{\Gamma}_R$,
$\hat{G}^{r}_{D,A}(\epsilon)$ and $\hat{G}^{a}_{D,A}(\epsilon)$
are constructed by coding to numerically calculate
$T_{LR}(\epsilon)$.

\end{document}